\documentclass{article}

\usepackage{microtype}
\usepackage{graphicx}
\usepackage{subcaption}
\usepackage{booktabs} 

\usepackage{hyperref}
\usepackage{multirow}

\usepackage[accepted]{icml2025}

\usepackage{amsmath}
\usepackage{amssymb}
\usepackage{mathtools}
\usepackage{amsthm}

\usepackage[capitalize,noabbrev]{cleveref}

\theoremstyle{plain}

\theoremstyle{definition}

\theoremstyle{remark}

\usepackage[textsize=tiny]{todonotes}

\icmltitlerunning{Temporal Representation Learning for Real-Time Ultrasound Analysis}

\begin{document}

\twocolumn[
\icmltitle{Temporal Representation Learning for Ultrasound Analysis using Masked Modeling}



\icmlsetsymbol{equal}{*}

\begin{icmlauthorlist}
\icmlauthor{Yves Stebler}{yyy}
\icmlauthor{Thomas M. Sutter}{equal,yyy}
\icmlauthor{Ece Ozkan}{equal,yyy,comp}
\icmlauthor{Julia E. Vogt}{yyy}
\end{icmlauthorlist}

\icmlaffiliation{yyy}{Department of Computer Science, ETH Zurich, Switzerland}
\icmlaffiliation{comp}{Department of Biomedical Engineering, University of Basel, Switzerland}

\icmlcorrespondingauthor{Ece Ozkan}{ece.oezkanelsen@unibas.ch}

\icmlkeywords{Machine Learning, ICML}

\vskip 0.3in
]



\printAffiliationsAndNotice{\icmlEqualContribution} 

\begin{abstract}
Ultrasound (US) imaging is a critical tool in medical diagnostics, offering real-time visualization of physiological processes. 
One of its major advantages is its ability to capture temporal dynamics, which is essential for assessing motion patterns in applications such as cardiac monitoring, fetal development, and vascular imaging. 
Despite its importance, current deep learning models often overlook the temporal continuity of ultrasound sequences, analyzing frames independently and missing key temporal dependencies. 
To address this gap, we propose a method for learning effective temporal representations from ultrasound videos, with a focus on echocardiography-based ejection fraction (EF) estimation. 
EF prediction serves as an ideal case study to demonstrate the necessity of temporal learning, as it requires capturing the rhythmic contraction and relaxation of the heart. 
Our approach leverages temporally consistent masking and contrastive learning to enforce temporal coherence across video frames, enhancing the model's ability to represent motion patterns. 
Evaluated on the EchoNet-Dynamic dataset, our method achieves a substantial improvement in EF prediction accuracy, highlighting the importance of temporally-aware representation learning for real-time ultrasound analysis.

\end{abstract}

\section{Introduction}
Ultrasound (US) imaging is one of the most widely used diagnostic tools in medicine due to its non-invasive nature, real-time feedback, and cost-effectiveness. 
It enables clinicians to visualize internal structures dynamically, making it indispensable for a wide range of medical applications, including obstetrics, cardiology, and emergency care \citep{Jensen2007, Edler2004}. 
One of the primary advantages of ultrasound is its ability to capture temporal information—sequential images over time that reflect physiological processes in real-time. 
This temporal aspect is crucial for assessing organ motion, blood flow, and dynamic physiological events.
These applications rely heavily on understanding motion and changes over time, which are not easily captured by static image-based models \citep{Thomas2006}. 
To fully harness the temporal richness of ultrasound, it is critical to learn effective temporal representations that encode motion patterns and sequential dependencies. 

Recent advancements in self-supervised learning have introduced Masked Autoencoders (MAEs), which have demonstrated strong capabilities in learning spatial representations by reconstructing masked input data \citep{He2022}. 
MAEs achieve this by randomly masking portions of an image and training the model to predict the missing parts, effectively learning rich feature representations in an unsupervised manner. 
However, their application has largely been restricted to frame-level analysis, where each frame is treated as an independent sample. 
This approach overlooks the sequential and continuous nature of ultrasound imaging, where physiological changes evolve smoothly over time. 
To extend MAEs for video understanding, VideoMAE was recently introduced, applying a similar masked reconstruction concept but optimized for video sequences, allowing for temporal feature extraction \citep{Tong2022}. 
While VideoMAE improves temporal learning over naive frame-based methods, its current implementations do not fully exploit the unique temporal dynamics of medical video sequences like ultrasound.
As a result, these models are limited in their ability to capture temporal dependencies critical for real-time assessment and clinical decision-making \citep{Bertasius2021}.
See \cref{sec:related_work} for more discussion on related work.

\begin{figure*}[ht]
    \centering
    \begin{subfigure}[t]{0.42\textwidth}
        \centering
        \includegraphics[width=\linewidth]{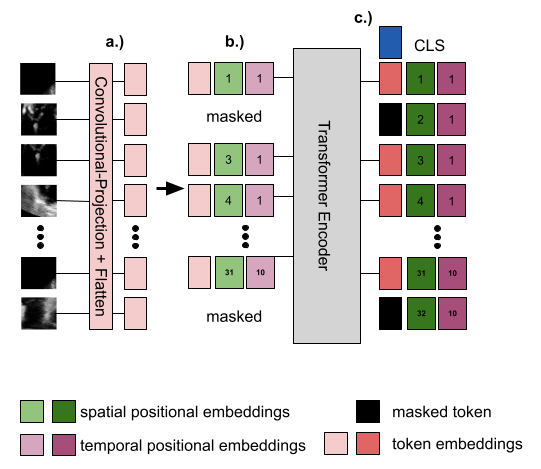}
        \caption{Encoder of the Temporal Model.}
        \label{fig:encoder}
    \end{subfigure}
    \hspace{0.05\columnwidth} 
    \begin{subfigure}[t]{0.42\textwidth}
        \centering
        \includegraphics[width=\linewidth]{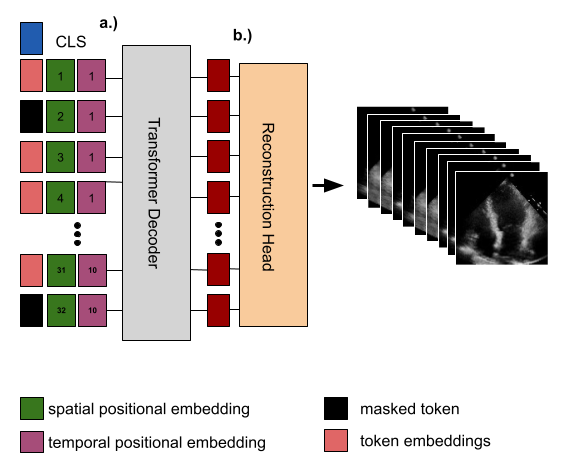}
        \caption{Decoder of the Temporal Model.}
        \label{fig:decoder}
    \end{subfigure}
    
    \caption{Overview of the Temporal Model. (i) The encoder extracts patches from the input frames, flattens them, and applies a learned spatial and temporal positional embedding to the unmasked patches, while removing masked patches from the sequence. (ii) The decoder reconstructs the original video by filling in masked tokens, reapplying positional embeddings, and passing through a transformer-based reconstruction process.}
    \label{fig:MAE_enc_dec}
\vspace{-0.5cm}
\end{figure*}

A prime example of the importance of temporal learning in ultrasound is echocardiography, where the goal is to measure cardiac function by analyzing sequences of heartbeats \citep{Ouyang2020, Zhang2018}. 
In particular, Ejection Fraction (EF) estimation, quantifying the percentage of blood ejected from the ventricles during each heartbeat—serves as a critical marker for heart health. 
Accurate EF prediction demands an understanding of the heart's motion across multiple frames, highlighting the need for effective temporal representation learning.

In this work, we specifically evaluate EF estimation as a case study to showcase the importance of temporal modeling in ultrasound imaging. 
Our method extends the MAE framework by incorporating temporally-aware mechanisms that enable the model to capture cardiac motion across sequences of frames. 
By learning temporally-aware representations, our approach significantly improves EF prediction accuracy, demonstrating the potential of temporal representation learning for real-time ultrasound analysis.

\begin{table*}[h!]
    \centering
    \footnotesize
    \caption{Binary classification results comparing frame-based and temporal models. The temporal model achieves better AUROC and recall, indicating stronger temporal feature capture.}
    \label{tab:binary_results}
    \begin{tabular}{llccccccc}
        \toprule
        \textbf{Model Type} & \textbf{Training Mode} & \textbf{Resolution} & \textbf{Model} & \textbf{F1 Score} & \textbf{Recall} & \textbf{Precision} & \textbf{Accuracy} & \textbf{AUROC} \\
        \midrule
        \multirow{3}{*}{Frame-based} 
            & Base & $32\times 32$ & ViT-T & 0.87 & 0.90 & 0.85 & 0.80 & 0.79 \\
            & End-to-End & $32\times 32$ & ViT-T & 0.89 & 0.89 & 0.83 & 0.83 & 0.86 \\
            & End-to-End, Oracle & $32\times 32$ & ViT-T & 0.89 & 0.89 & 0.88 & 0.82 & 0.84 \\
        \midrule
        \multirow{4}{*}{Temporal} 
            & Base & $32\times 32$ & ViT-T & 0.88 & 0.85 & 0.92 & 0.80 & 0.77 \\
            & End-to-End & $32\times 32$ & ViT-T & 0.89 & 0.87 & 0.90 & 0.82 & 0.83 \\
            & End-to-End, Oracle & $32\times 32$ & ViT-T & 0.89 & 0.85 & 0.93 & 0.82 & 0.82 \\
            & End-to-End, Contrastive, Oracle & $32\times 32$ & ViT-T & 0.89 & 0.87 & 0.92 & 0.84 & 0.88 \\
        \bottomrule
    \end{tabular}
\vspace{-0.5cm}
\end{table*}

\section{Methods}
Our approach extends the original Masked Autoencoder (MAE) architecture \citep{He2022} to effectively capture temporal dynamics and spatial features from video-based ultrasound sequences.
The key novelty of our approach lies in how we handle temporal information during encoding, allowing the model to learn not just spatial features, but also temporal dynamics critical for medical video analysis.
\renewcommand{\thesubfigure}{\roman{subfigure}}

\subsection{Preprocessing and Video-Based Token Extraction}
To efficiently encode temporal data, we preprocess the ultrasound video by stacking frames sequentially and dividing each frame into non-overlapping patches. 
Specifically, each video frame $X_t$ is processed independently, ensuring non-overlapping spatial regions. 
This results in a set of spatial patches $P_t$ for frame $t \in [1, T]$, where $T$ is the number of frames. 
These patches are then flattened and linearly embedded into the encoder's latent space, preserving the spatial and temporal structure across frames. 
The entire video sequence is represented as a flattened tensor of shape $X_{\text{flattened}} \in \mathbb{R}^{(T \cdot N) \times D}$, where $N$ is the total number of patches per frame, and $D$ is the embedding size. 
This structured representation allows the transformer to learn dependencies both within individual frames and across the temporal axis.

To explicitly encode temporal relationships, we introduce \textit{temporal-positional embeddings} that encode both the frame order and patch positions as $E_t = E_{\text{pos}}(t) + E_{\text{time}}(t)$, where $E_{\text{pos}}(t)$ represents the spatial position within the frame and $E_{\text{time}}(t)$ encodes the temporal sequence across frames.

\subsection{Frame-wise Random Masking}  
Unlike traditional tube-like masking strategies \citep{Tong2022, Kim2024}, we adopt a \textit{frame-wise random masking strategy}. 
For each frame $t$, we randomly select a subset of patches to be masked
\begin{equation}
M_t = \text{Mask}_t (P_t), \quad \forall t \in [1, T].
\vspace{-0.5cm}
\end{equation}

The randomness in masking is applied independently across frames $t$, ensuring that different spatial regions are masked over time. 
This design encourages the model to reconstruct not only the missing patches but also the motion dynamics that link frames, reinforcing spatiotemporal learning during pretraining.

\subsection{Pretraining Objective}  
Our pretraining process optimizes two complementary objectives. 

\paragraph{Reconstruction Loss:}  
The reconstruction loss is applied over the masked patches, compelling the model to restore the original spatial details with

\begin{equation}
L_{\text{rec}} = \frac{1}{|M|} \sum_{t=1}^T \sum_{i \in M} \| \hat{P_t}[i] - P_t[i] \|^2,
\end{equation}
where $P_t[i]$ is the original patch at index $i$ in frame $t$, $\hat{P_t}[i]$ is its reconstructed version, and $M$ is the set of masked patches.
This loss encourages high-fidelity reconstruction of local image features while learning robust spatial representations.

\paragraph{Temporal Contrastive Loss:}  
To capture temporal coherence, we compute frame-level representations as the average of all patch tokens in each frame as
\begin{equation}
f_t = \frac{1}{N} \sum_{i=1}^N p_t[i] \in \mathbb{R}^D.
\end{equation}

For any pair of frames $(t, t + \Delta t)$, we calculate their cosine similarity
\begin{equation}
\cos(f_t, f_{t+\Delta t}) = \frac{f_t \cdot f_{t+\Delta t}}{\|f_t\| \|f_{t+\Delta t}\|}.
\end{equation}

We further derive the cosine distance as
\begin{equation}
d_{t, \Delta t} = 1 - \cos(f_t, f_{t+\Delta t}).
\end{equation}

The contrastive loss encourages temporally close frames to have similar representations while enforcing a margin for distant frames with
\begin{equation}
L_{\text{contrast}} = \frac{1}{C} \sum_{t=1}^T \sum_{\Delta t=1}^{T - t}
\begin{cases} 
  d_{t, \Delta t}^2 & \text{if } \Delta t \leq \tau_p \\
  [\tau_m - d_{t, \Delta t}]_+^2 & \text{if } \Delta t > \tau_p
\end{cases}
\end{equation}

where $\tau_p$ is the threshold for positive temporal consistency, $\tau_m$ is the margin for negative temporal separation and $C$ is the total number of temporal comparisons, where 
\begin{equation}
C = \sum_{t=1}^T (T - t).
\end{equation}

The overall pretraining objective is a weighted sum of the reconstruction and contrastive losses
\begin{equation}
L_{\text{total}} = L_{\text{rec}} + \lambda L_{\text{contrast}},
\end{equation}

where $\lambda \in [0, 1]$ balances the influence of spatial reconstruction and temporal alignment.

\subsection{Downstream Tasks}
For downstream tasks, we utilize the encoded \textit{CLS token} from the final transformer block in the encoder as the input to a lightweight regression head. 
This CLS token, enriched with temporal and spatial representations, serves as a summary of the video sequence, enabling high accuracy in clinical predictions.

\begin{table*}[h!]
\centering
\footnotesize
\caption{Binary classification results from \citet{Zhang2024} compared with our temporal contrastive loss model.}
\label{tab:VideovsFrameResultsBinary_Contrastive_Compare}
\begin{tabular}{lccccccc}
\toprule
\textbf{Model} & \textbf{Dataset} & \textbf{Params} & \textbf{Video-Input} & \textbf{F1} & \textbf{Accuracy} & \textbf{AUROC} \\
\midrule
VideoMAE & $\sim$200,000 & $\sim$98M & $16\times224\times224$ & 0.92 & 0.88 & 0.91 \\
ECHO-VISION-FM & $\sim$200,000 & $\sim$98M & $16\times224\times224$ & 0.93 & 0.89 & 0.93 \\
Ours (Temporal, End-to-end, Oracle) & $\sim$10,000 & $\sim$8M & $10\times32\times32$ & 0.89 & 0.84 & 0.88 \\
\bottomrule
\end{tabular}
\vspace{-0.5cm}
\end{table*}

\section{Experiments and Results}
\subsection{Experimental Setup}
Our experimental setup is designed to evaluate the effectiveness of temporal representation learning in real-time ultrasound video analysis, specifically targeting Ejection Fraction (EF) estimation. 
All experiments are conducted on the EchoNet-Dynamic dataset \citep{Ouyang2020}, which comprises approximately 10,000 echocardiogram videos, each annotated with EF values.

We employ a temporal backbone model that processes sequences of 10 frames per input video, with each frame downsampled to $32\times32$ resolution for computational reasons.
The backbone follows the Vision Transformer (ViT) architecture, using the ViT-Tiny and ViT-Base variants as proposed by \citep{Dosovitskiy2020,Wu2022}.
The frames are uniformly sampled over a one-second interval of the cardiac cycle, ensuring the capture of critical moments such as End Diastolic Volume (EDV) and End Systolic Volume (ESV). 
This sampling strategy is intended to preserve the temporal structure of heart motion for better feature extraction.

Pretraining is performed using a combination of a masked reconstruction objective and our proposed temporal contrastive loss, which encourages the model to learn both spatial and temporal representations effectively.
The model is trained using the AdamW optimizer with a base learning rate of $1.5 \times 10^{-4}$, adjusted by the batch size, and a weight decay of $0.05$. 
The learning rate schedule is managed by a LambdaLR scheduler, which applies a warm-up period during the first 200 epochs, followed by a cosine decay. 
We implement early stopping with a patience threshold of 75 epochs, terminating the training if the reconstruction loss does not improve by at least $5\times10^{-5}$.

For the downstream classification task, we perform binary classification to distinguish between normal (EF $> 50\%$) and reduced EF (EF $\leq 50\%$). 
The classification head consists of two fully connected layers of sizes 256 and 128, respectively, and uses the CLS token output from the encoder transformer. 
The model is evaluated using standard classification metrics, including F1 Score, Recall, Precision, Accuracy, and AUROC.

\paragraph{Training Configurations}
We evaluate following training configurations:

\textit{Base Training}: In this setting, the encoder is frozen and only the classification head is trained. 
This setup serves as a lower-bound baseline to isolate the quality of the learned representations without further fine-tuning.

\textit{End-to-End Training}: In this configuration, both the encoder and the classification head are jointly optimized during the fine-tuning phase. 
This allows for simultaneous gradient updates across the entire architecture, enhancing feature extraction and classification alignment.

\textit{Contrastive Training}: This mode uses our temporal contrastive loss during pretraining to encourage temporal consistency in learned representations, complementing the spatial reconstruction objective.

\textit{Oracle Setting}: This setup assumes optimal frame selection during pretraining and inference, where frames are perfectly aligned with key cardiac phases, such as systole and diastole. 
This configuration serves as an upper bound on achievable performance, providing insight into the maximum potential of our temporal modeling approach.

\textit{Contrastive + Oracle} combines both the contrastive pretraining and oracle-aligned input, reflecting the best-case temporal modeling performance.

\subsection{Binary Classification Results}
\Cref{tab:binary_results} summarizes the performance of all evaluated models and training configurations for the binary classification task. 
Notably, the temporal model trained with our contrastive loss under the oracle setting achieves the highest AUROC of 0.88, outperforming all frame-based counterparts, including those trained end-to-end. 
This result underscores the importance of modeling temporal dependencies explicitly, as the contrastive objective effectively encourages the model to capture the motion dynamics across frames.

\subsection{Comparison with State-of-the-Art}
We compare our model with the work of \citet{Zhang2024}, as shown in \Cref{tab:VideovsFrameResultsBinary_Contrastive_Compare}. 
For pretraining, the authors utilized 40\% of the MIMIC-IV-ECHO dataset \citep{mimic-iv-echo}, which comprises approximately 500,000 echocardiogram videos. 
Their classifier was fine-tuned on the EchoNet-Dynamic dataset. 
They also employed an input resolution of $224\times224$ and processed 16 frames per forward pass using a ViT-B backbone.

Although specific parameter counts are not provided in their work, the authors mention that their configuration closely follows the original ViT-B/16 VideoMAE design, on which our parameter estimation is based. 
Despite using significantly less training data, a smaller model size, lower input resolution, and fewer frames, our model achieves competitive performance with ECHO-Vision-FM.

\section{Discussion}
Our experimental results demonstrate that the proposed temporal MAE-based model effectively captures the temporal dynamics of cardiac cycles, outperforming frame-based baselines in EF estimation. 
By leveraging ViT-Tiny and ViT-Base backbones, our model achieves competitive performance with state-of-the-art methods while requiring significantly less data and computational resources. 
The introduction of the Temporal Contrastive Loss further enhances the temporal consistency of learned representations, contributing to improved classification accuracy. 
Although our model shows promising results, future work could explore adaptive frame selection and multi-view echocardiography to further enhance temporal feature extraction. 
Overall, our findings underscore the importance of temporally-aware self-supervised learning for real-time ultrasound analysis.

\bibliography{ref.bib}
\bibliographystyle{icml2025}

\appendix

\section{Related Work}
\label{sec:related_work}
Research in ultrasound representation learning has explored a range of methods to improve downstream task performance. 
Traditional approaches predominantly focused on frame-based learning, where spatial features are extracted independently from each frame without consideration of temporal continuity. 
More recent works, however, have shifted towards learning representations that incorporate temporal dynamics, recognizing the importance of capturing motion and periodicity in ultrasound sequences.

\paragraph{Ultrasound Representation Learning}
Recent works have demonstrated the advantages of learning robust representations from ultrasound images for downstream tasks. 
\citet{Droste2019} introduced a self-supervised learning method that pretrains a CNN using a visual-tracking dataset to predict saliency maps corresponding to sonographer focus points. 
Pretraining with visual-tracking information improved F1 performance on a standard plane-detection task, outperforming direct fine-tuning, highlighting the importance of ultrasound-specific feature learning. 
Similarly, \citet{Zeyu2022} proposed Anatomy-Aware Contrastive Learning for self-supervised ultrasound representation learning. 
By grouping positive pairs based on anatomical similarity, their method effectively captured granular information, improving performance in standard plane classification and fetal biometry estimation compared to ImageNet-pretrained models.

\paragraph{Temporal Representation Learning}
While frame-based methods have shown promising results, recent research emphasizes the importance of modeling temporal dynamics for tasks like cardiac monitoring and fetal development analysis. 
Authors in \citep{Jiao2020} introduced a self-supervised framework for ultrasound video representation learning that leverages video-frame sequence ordering and geometric transformation as pretext tasks. 
This approach effectively captured temporal dependencies, outperforming static frame-based baselines. 
\citet{Zhang2024} proposed ECHO-Vision-FM, which combined VideoMAE with a Spatio-Temporal Feature Fusion Network (STFF-Net). 
VideoMAE, serving as the backbone, utilized tube-masking strategies to maintain temporal consistency across frames, significantly improving performance on EF prediction tasks. 
Authors in \citep{Kim2024} extended this concept with EchoFM, introducing a periodic contrastive loss that enforces temporal consistency within cardiac cycles. 
This strategy enhanced the learning of periodic heart motion, resulting in improved segmentation and EF prediction accuracy.

\end{document}